# Design considerations and performance analysis of fiber laser array system for structuring orbital angular momentum beams

TIANYUE HOU,[1] QI CHANG,[1] JINHU LONG,[1] PENGFEI MA,[1, 2] AND PU ZHOU[1, 3]

[1]*College of Advanced Interdisciplinary Studies, National University of Defense Technology, Changsha 410073, China*
[2]*shandapengfei@126.com*
[3]*zhoupu203@163.com*

**Abstract:** Since the advent of optical orbital angular momentum (OAM), advances in the generation and manipulation of OAM beams have continuously impacted on intriguing applications including optical communication, optical tweezers, and remote sensing. To realize the generation of high-power and fast switchable OAM beams, coherent combining of fiber lasers offers a promising way. Here in this contribution, we comprehensively investigate the coherent fiber laser array system for structuring OAM beams in terms of the design considerations and performance analysis. The performance metric and evaluation method of the laser array system are presented and introduced. Accordingly, the effect of the main sections of the laser array system, namely the high-power laser sources, emitting array configuration, and dynamic control system, on the performance of the output coherently combined OAM beams is evaluated, which reveals the system tolerance of perturbative factors and provides the guidance on system design and optimization. This work could provide beneficial reference on the practical implementation of spatially structuring high-power, fast switchable OAM beams with fiber laser arrays.

## 1. Introduction

Owing to the exotic intensity and phase spatial structure as well as the unique dynamics characteristic, optical vortex beams carrying orbital angular momentum (OAM) have attracted substantial research interests [1-4]. As a physical dimension, the OAM of light is theoretically unbounded, which not only offers a promising platform for advanced scientific research, but also spurs the developments in widespread applications including the optical communication [5-8], particle manipulation [9-11], optical imaging [12,13], and remote sensing [14, 15]. The growing interest in the fundamental research and applications the OAM dimension have impelled calls for the intense attention focused on the efficient generation of OAM-carrying optical vortex beams. Up to now, a fair amount of approaches to generate OAM beams have been put forward and demonstrated [6,16]. Specifically, one feasible way that has been well investigated is to convert the Gaussian beam into an OAM beam via using an external-cavity converter, which can be the spatial light modulator [17], spiral phase plate [18] and slit [19], metamaterial [20], and grating [21,22]. To overcome the power loss and enhance the modal purity, tailoring OAM beams at the laser source with special intracavity design, as another promising pathway, has been significantly developed [23-26]. Despite the impressive progress of OAM beams generation, there has been an ongoing challenge that limits the practical usefulness of OAM beams, namely creating high-power OAM beams and realizing high switching rate of the OAM mode at the same time. The challenge for the generation of high-power and fast switchable OAM beams has stimulated the recent development of another possible solution, namely spatially structuring OAM beams with laser arrays based on coherent beam combining (CBC) technology [27].

   Historically, CBC of fiber lasers has been celebrated for its utility in scaling the output power of laser beam while maintaining good beam quality [28-31]. In the recent years, tailoring the spatial profile of light based on CBC systems has become a reality [32-35]. By harnessing the active phase control of array elements, the laser beams can mutually interfere and "twist" the phase front of combined beam to form the optical vortex in the far-field. For a coherent fiber laser array system, the inherent power scaling capacity of CBC technology holds great promise for the power enhancement of the output OAM beams, and the characteristics of devices and control circuit to implement phase modulation suggest the agility of OAM mode switching [36]. Driven by these advantages, substantial efforts have been made to construct the theoretical model of structuring OAM beams from coherent laser arrays, in terms of optical field formation [37,38], propagation properties [39], phase stabilization [36,40], and OAM mode detection and characterization [27,41]. Meanwhile, experimental demonstrations [32,34,42-45] for the laser array based generation of OAM beams have been carried out. The theoretical and experimental achievements preliminarily verified the feasibility of the technical route. Despite the exciting achievements, it is worth mentioning that the combining efficiency of laser arrays and modal purity of combined beam depend on the characteristics of the laser sources, emitting array configuration, and control system, as the main sections of the laser array system. When the system is operated in perturbative states or without elaborate design, the system performance suffers. Up to this point, the performance evaluation and design consideration of the fiber laser array systems for tailoring OAM beams thus becomes the most vital step towards the further development of the technical route.

   In this paper, we explore the performance evaluation and design considerations of the coherent fiber laser array system for structuring OAM beams. We first illustrate that the routinely utilized performance metrics for the optical vortices and conventional CBC systems are inappropriate to evaluate the performance of the laser array system that customizes OAM beams. Then, according to the theoretical concept of higher-order Airy patterns [27], the performance metric and general evaluation method of the laser array system to tailor OAM beams are put forward and introduced. Based on the proposed metric and evaluation method, the requirements and key factors of the high-power laser sources, emitting array configuration, and dynamic control system the performance of the coherently combined OAM beams depends on are analyzed, for the laser array system, which could guide the design and optimization for the configuration of system. This work offers valuable guidance and comprehensive references on the evaluation, design, and further development for the generation of high-power, fast switchable OAM beams with coherent fiber laser arrays.

## 2. System Model

### 2.1 Architecture of laser array system

Consider a coherent fiber laser array system for structuring OAM beams, as is schematically depicted in Fig. 1. The typical laser array system consists of three main sections: (i) the high-power laser sources that generate and amplify the fiber laser beams of multiple channels, (ii) the emitting array configuration that collimates, arranges, and emits the laser beams to free space, and (iii) the dynamic control system that implements real-time wavefront control to the array element of each channel to compensate the aberrations and realize the flexible OAM mode shifting. Here we introduce how the three main sections operate in cooperation to ensure the controllable generation of OAM beams from the laser array system.

In the high-power laser sources section, the output of seed laser (SL) transmits through a pre-amplifier (PA) to preliminarily scale the power. Then, the power scaled laser beam is split into multiple channels via a fiber splitter (FS). As for each channel, the laser beam is modulated by a phase modulator (PM) and subsequently propagates through cascaded fiber amplifiers (CFAs) for further power enhancement. The power scaled laser beams, as the output of the high-power laser sources section, are sent to the emitting laser array configuration section. The emitting laser array configuration determines the output formatting of the laser array, which serves as an important basis for spatial beam shaping. The emitting laser array configuration can be the adaptive fiber optics collimator (AFOC) array, which is composed of fiber positioner devices and collimating lens [46]. It has been demonstrated that utilizing the AFOC configuration, highly efficient beam projection and tip-tilt wavefront control could be achieved at the same time [47-49]. Based on the emitting laser array configuration, the collimated laser beams are tiled side by side and propagate to free space at the source plane.

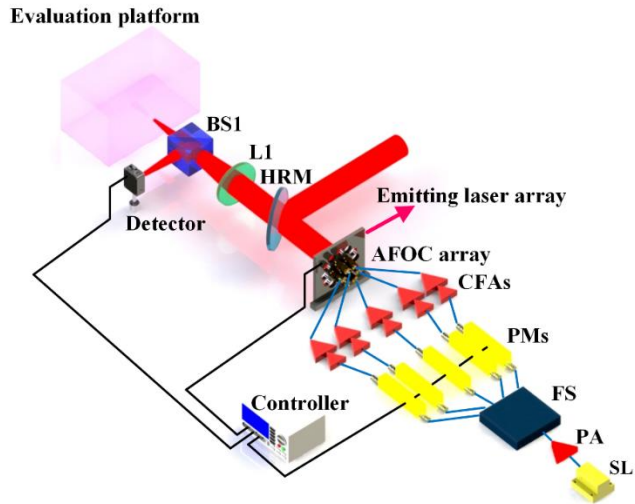

Fig. 1. Architecture of coherent fiber laser array system for structuring OAM beams. (SL: seed laser; PA: pre-amplifier; FS: fiber splitter; PM: phase modulator; CFAs: cascaded fiber amplifiers; AFOC: adaptive fiber optics collimator; HRM: high reflective mirror; L: focus lens; BS: beam splitter.)

The output laser beams of the emitting laser array configuration section are split by a high reflective mirror (HRM). The high-power reflected part propagates to the far-field and forms the optical vortex beam by the mutual interference, whereas the transmitted part of lower power is utilized for dynamic wavefront control and combining performance evaluation. Specifically, the lower-power part of laser beams that transmits the HRM is focused by a focal lens (L1) prior to being sampled by a beam splitter (BS1). Part of the focused laser beams transmits via the BS1 and is sent to the performance evaluation platform, of which more detailed information would be elucidated in section 2.4. The other part that is reflected by the BS1 is coupled into the detection module, which extracts the optical information of the coherently combined OAM beam, converts the optical information to electrical signals, and sends the signals to the dynamic control section. In the dynamic control section, the controller continuously processes the received signals by performing the piston phase and tip-tilt wavefront control algorithms. Correspondingly, the control voltages are generated by the controller and applied to the PMs and AFOC array to realize the closed-loop wavefront control. On the one hand, when the dynamic control section of laser array system is operated in closed loops, dynamic phase errors and tip-tilt wavefront distortions that are induced by thermal and environmental fluctuations can be compensated. On the other hand, active phase control voltages can be generated and applied to the PMs to enable the piston phase shifting, thus ensuring the flexible switching of OAM modes.

## 2.2 Mathematical characterization for laser array

To facilitate the analysis for the performance evaluation and system design, the mathematical characterization for the laser array, as a fundamental part of the system model, is necessary to be illustrated. In fact, the construction of the theoretical model for coherent laser array to tailor OAM beams has been addressed in previous publications. However, direct use of the published formula would cause ambiguousness and inconsistencies in numbering, symbols, and derivations, and moreover, systematical considerations for some of the characteristics of the laser array system are not covered. Hence, for convenience of following investigations and the benefit of the audience, we briefly introduce the mathematical characterization of laser array before analyzing the performance metrics and evaluation.

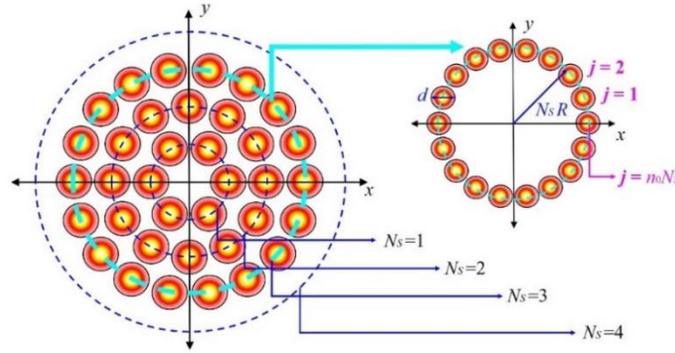

Fig. 2. Schematic setup for the emitting laser array of coherent fiber laser array system to tailor OAM beams.

Consider a $N$-element coherent fiber laser array system for structuring OAM beams, of which the emitting laser array is shown in Fig. 2. The emitting laser array at the source plane is composed of $n$ concentrically arranged radial subarrays, and for each subarray, the laser beams are tiled uniformly along the azimuthal direction. Here we define $N_S$ as the serial number of radial subarrays, and the $N_S$-th radial subarray could contain $n_0 N_S$ beamlets. The central position $(x_{j,N_S}, y_{j,N_S})$ of the $j$-th beamlet lying on the $N_S$-th radial subarray is given by

$$\begin{cases} x_{j,N_S} = N_S R \cos\left(\dfrac{2\pi j}{n_0 N_S}\right) \\ y_{j,N_S} = N_S R \sin\left(\dfrac{2\pi j}{n_0 N_S}\right) \end{cases}, \quad (1)$$

where $R$ denotes the structure parameter that determines the compactness of the laser array. As for the $N$-element laser array, the serial numbers of the inner subarray and outer subarray are assumed to be $n_1$ and $n_2$, respectively. We assume that each laser beam is truncated by a circular aperture with the diameter of $d$, and accordingly, the diameter of the entire aperture for the laser array yields $D = 2n_2 R + d$. To generate the OAM beam with the topological charge (TC) of $m$, the mathematical form of the optical field at the source plane is expressed as

$$\begin{aligned} \mathbf{E}(x,y) = \sum_{n=n_1}^{n_2} \sum_{j=1}^{nn_0} A_{j,n} &\exp\left[-\dfrac{(x-x_{j,n})^2 + (y-y_{j,n})^2}{w_0^2}\right] \\ &\times circ\left[\dfrac{\sqrt{(x-x_{j,n})^2 + (y-y_{j,n})^2}}{(d/2)}\right] \exp\left(im\dfrac{2\pi j}{nn_0}\right)\exp(i\Delta\phi_{j,n}) \\ &\times \exp\left[i\dfrac{2\pi(x\sin\Delta\alpha_{j,n} + y\sin\Delta\beta_{j,n})}{\lambda}\right]\left(\mathbf{e_x}\cos\chi_{j,n} + \mathbf{e_y}\sin\chi_{j,n}\right) \end{aligned}, \quad (2)$$

where $circ(r_0)$ denotes the transmittance function of the circular aperture, viz.,

$$circ(r_0) = \begin{cases} 1 & r_0 < 1 \\ 1/2 & r_0 = 1 \\ 0 & otherwise \end{cases}. \tag{3}$$

$\lambda$, $w_0$, and $A_{j,n}$ account for the wavelength, waist width, and amplitude of the $j$-th beamlet lying on the $n$-th radial subarray, respectively. $\mathbf{e_x}$ and $\mathbf{e_y}$ represent the unit vectors of the $x$ and $y$ axes, respectively. We assume that the desirable polarization state is linearly along $\mathbf{e_x}$. $\Delta\phi_{j,n}$, $(\Delta\alpha_{j,n}, \Delta\beta_{j,n})$, and $\chi_{j,n}$ are, in turn, the piston phase error, tilt error angle, and the depolarization angle from the desirable polarization state. In this work, we focus our attention on the generation of scalar optical vortex beams, thus the desirable polarization optical field component $E_x(x, y)\mathbf{e_x}$ would be further discussed.

In the free space, the emitted laser beams propagate and the optical fields are coherently superposed. Based on the paraxial diffraction theory, the combined OAM beam at the receiver plane can be expressed as [50]

$$E_c(x, y, z = L) = \frac{e^{ikL}}{i\lambda L} \int_{-\infty}^{\infty}\int_{-\infty}^{\infty} E_x(u, v, z = 0) e^{\frac{ik}{2L}\left[(x-u)^2 + (y-v)^2\right]} du dv, \tag{4}$$

where $k = 2\pi/\lambda$ is the wave number, and $L$ denotes the distance between the source plane and receiver plane. Accordingly, the intensity distribution of the coherently combined beam yields

$$I_c(x, y, z = L) = \left[E_c(x, y, z = L)\right]\left[E_c(x, y, z = L)\right]^*. \tag{5}$$

In the following sections, the theoretical model and mathematical characterization of laser array we present here are employed to analyze the system performance.

## 2.3 Inappropriateness of conventional metrics

To evaluate the performance of the coherent fiber laser array systems for structuring OAM beams, there is an intuitive idea that utilizing the existing beam quality metrics of laser beams. The Strehl ratio and $M^2$, as widely known laser beam quality metrics, have been extensively used for the performance evaluation of laser systems [51,52]. In spite of the commonality, the Strehl ratio, which is defined as the ratio of the peak, on-axis intensity to that of an ideal flat-top, filled aperture reference beam, cannot characterize the beam quality of the combined OAM beams, since the OAM beams possess doughnut-shaped intensity profiles and null on-axis intensity attributed to the helical phase structures that carry optical vortex cores. The $M^2$-parameter is defined as the ratio of the product of divergence angle and spot size for the measured beam to that of a fundamental Gaussian mode reference beam. The propagation properties of optical vortex beams are quite different form the fundamental Gaussian mode, thus the $M^2$-parameter that reveals the similarity of the measured beam and ideal fundamental Gaussian mode is inappropriate to evaluate the beam quality of the combined OAM beams. One may well ask that whether it is feasible to utilize a modified $M^2$-parameter as a performance metric by extending the reference beam from the fundamental Gaussian mode to the set of Laguerre–Gaussian modes. In fact, to acquire the $M^2$-parameter and its extended metrics, the correct calculation or measurement of the second-moment beam radius is necessary. However, as for the coherently combined beams generated from laser arrays, it has been demonstrated that the sidelobes of relative low intensity formed by the interference of beamlets would excessively magnify the magnitude of the second-moment beam radius, thus it is difficult to correctly characterize the beam quality of combined OAM beam by using the $M^2$ and its related metrics [52].

In previous studies, the power-in-the-bucket (PIB) and its related metric beam propagation factor (BPF) have been verified to be useful for evaluating the performance of the combined beams generated by laser arrays [49,53-55]. The PIB metric refers to the integral of the intensity distribution encircled in an on-axis circular area of a specific radius at the receiver plane, namely,

$$PIB = \int_0^{2\pi}\int_0^{r_{PIB}} I_c(\rho, \psi, z = L)\rho d\rho d\psi. \tag{6}$$

where $(\rho, \psi)$ represent the *Polar* coordinates of the receiver plane, $r_{PIB}$ denotes the radius of the circular integral area, and $I_c(\rho, \psi, z = L)$ is the intensity distribution of the *Polar* coordinates representation at the receiver plane. To analyze the utility of PIB metric in characterizing the performance of the combined OAM beams, numerical simulation has been conducted, and the results are displayed in Fig. 3. Consider the laser array system composed of 2 subarrays and 18 beamlets to tailor the OAM beam with a TC of 1 as an example. The intensity profile of the laser array at the source plane and the integral area to calculate the PIB metric are depicted in Figs. 3(a) and 3(b), respectively. For the convenience of calculation, a focus lens with a focal length of $f = 20$ m is assumed to be positioned behind the laser array, and accordingly, the intensity distribution of the combined beam in the far-field can be equivalently obtained at the Fourier plane. The radius of the integral area $r_{PIB}$ is set to be 0.2896 mm, which ensures the bucket to contain the main ring of the coherently combined OAM beam. Figures 3(c1)-3(c3) exhibit the phase distributions of the laser array to generate the OAM beams with TC = 1, TC = -1, and TC = 2, respectively, and the corresponding far-field intensity distributions are presented, in turn, in Figs. 3(d1)-3(d3). In terms of the definition in Eq. (6), the integral area, and the intensity distributions, the normalized PIB for the cases of generating the OAM beams with TC = 1, TC = -1, and TC = 2 are calculated as 0.5110, 0.5110, and 0.3970, respectively. The PIB metric can indeed reveal the energy concentration of the combined beam, and characterize the combining efficiency to some extent. However, we can see that the properties of the combined beam that associates to the phase structure are entirely omitted. Hence, we can conclude that directly using the PIB metric cannot characterize the performance of combined OAM beams accurately.

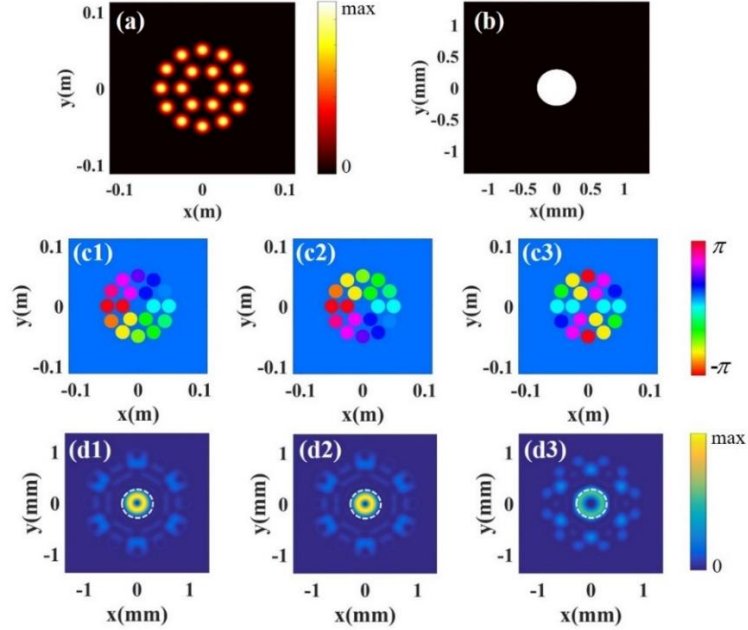

Fig. 3. Results of using PIB metric to characterize combined OAM beams. (a) Intensity profile of the 18-element laser array. (b) Integral area of the "bucket" in the far-field. (c1), (c2), and (c3) are the phase distributions of the laser array to generate OAM+1, OAM-1, and OAM+2 beams, respectively. (d1), (d2), and (d3) are the far-field intensity distributions of the combined OAM+1, OAM-1, and OAM+2 beams, respectively.

When compared to the PIB metric, OAM mode purity and OAM spectrum can reflect both the intensity profile and phase structure, which are routinely used to characterize OAM beams [56,57]. According to the definition of OAM mode purity and OAM spectrum, the optical field of coherent laser beam can be represented by the composition of spiral harmonics. The weights of power correspond to the spiral harmonics constitute the OAM spectrum, and the relative intensity of the harmonic $\exp(il\psi)$ is defined as the purity of the $l$-th OAM mode. Here, we attempt to analyze the performance of coherently combined OAM beams by using the OAM mode purity as the criterion. We firstly consider a 36-element laser array that consists of 3 subarrays for generating OAM+1 beams. The intensity and phase distributions of the laser array at the source plane, and the far-field intensity distribution of the combined OAM beam are shown in Figs. 4(a1), 4(b1), and 4(c1), respectively. Based on numerical calculations, the OAM spectrum of the combined OAM beam is obtained and displayed in Fig. 4(d1). In comparison, we take an 18-element laser array for generating OAM+1 beams as an example, of which the intensity profile and phase distribution at the source plane are shown in Figs. 4(a2) and 4(b2), respectively. The far-field intensity distribution of the combined beam is shown in Fig. 4(c2), and Fig. 4(d2) exhibits the numerically calculated OAM spectrum. The results indicate that for the two laser arrays, the purity of the OAM+1 mode is approximately the same (73.4% for the 36-element laser array, and 73.9% for the 18-element laser array). However, the coherent combination of the 36-element laser array forms the combined OAM beam with concentrated energy in the main ring, whereas the energy spreads from the main ring to the side rings for the 18-element laser array case. In spite of comprehensively considering both the intensity and phase information, the calculation lacks of using the radial information, resulting in the fact that the OAM mode purity metric cannot reveal the differences in the energy distribution along the radial axis. Therefore, directly using OAM mode purity as the performance metric is also improper to characterize the coherently combined OAM beams.

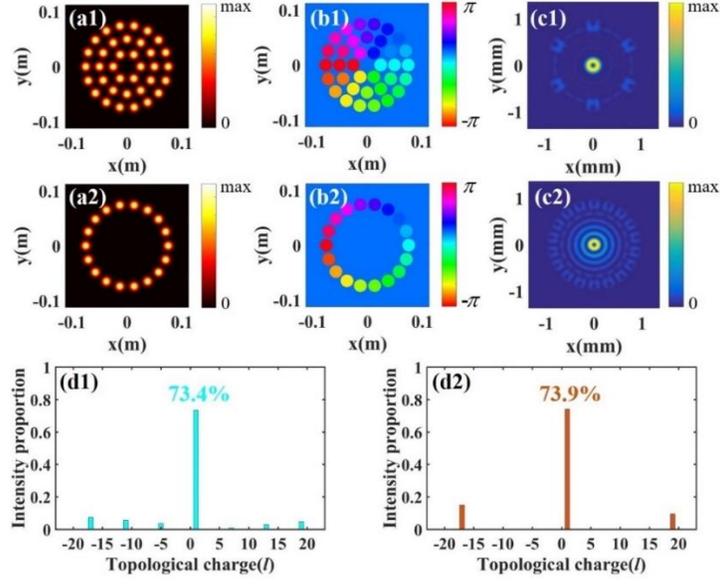

Fig. 4. Results of using OAM mode purity as performance metric to characterize combined OAM beams. (a1) Intensity profile and (b1) phase distribution of the 36-element laser array at the source plane to generate OAM+1 beam. (c1) Intensity distribution and (d1) OAM spectrum of the OAM+1 beam formed by the 36-element laser array. (a2) Intensity profile and (b2) phase distribution of the 18-element laser array at the source plane to generate OAM+1 beam. (c2) Intensity distribution and (d2) OAM spectrum of the OAM+1 beam formed by the 18-element laser array.

In a nutshell, conventional metrics for laser beam characterization including Strehl ratio, $M^2$-parameter, PIB, and OAM mode purity have been analyzed to evaluate the performance of combined OAM beams. The results indicate that each of the metrics can reflect partial characteristics of the combined beam, whereas evaluating the combining performance comprehensively and accurately is still challenging. Hence, we are motivated to propose an appropriate performance metric and evaluation method for the laser array system.

**2.4 Performance metric and evaluation method**

As we have discussed in the section 2.3, the conventional metrics that are routinely utilized to characterize the performance of a single laser beam, phase-locked laser arrays, and optical vortex beams cannot accurately reflect the performance of the coherent laser array system for structuring OAM beams. Hence, we are motivated to propose an appropriate performance metric and put forward a general method for the measurement and calculation of the metric to evaluate the system performance.

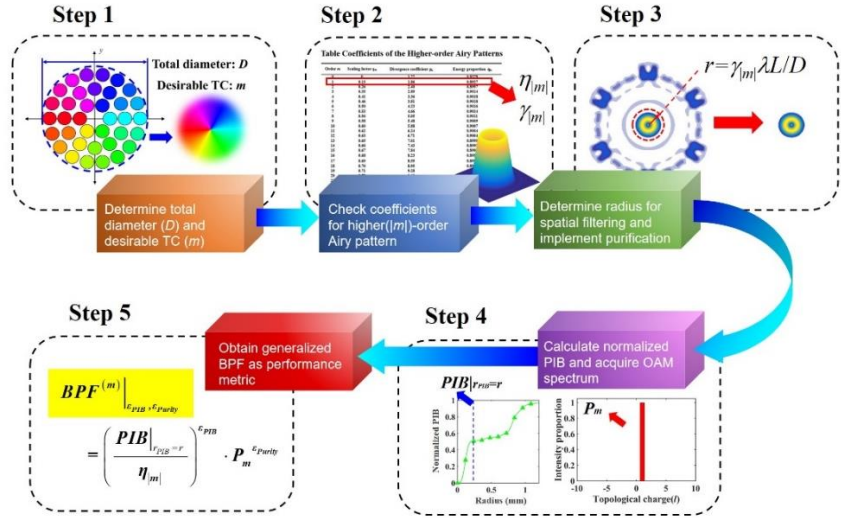

Fig. 5. Block diagram for the general method to acquire the performance metric and evaluate the coherent fiber laser array system for structuring OAM beams.

Figure 5 presents the block diagram for the general method to acquire the metric and evaluate the performance of the coherent fiber laser array system for structuring OAM beams. Firstly, according to the geometric arrangement and dimensions of the beamlets at the source plane, the total diameter $D$ of the entire aperture for the laser array can be determined by $D = 2n_2R + d$. Meanwhile, the desirable TC of the combined OAM beam $m$ related to the phase shifting of each beamlet is determined as a prior knowledge. Subsequently, referring to the concept and coefficients table of the higher-order Airy patterns [27], we can obtain the divergence coefficient $\gamma_{|m|}$ and energy proportion $\eta_{|m|}$ that correspond to the absolute value of index $m$. The concept of higher-order Airy pattern is extended from the fundamental Airy patterns (Fraunhofer diffraction pattern of the plane wave illuminated circular aperture), as a classic case of the diffraction theory [51,58], by introducing the OAM of light, which has been elucidated in our previous publication [27]. In the third step, beam purification of the combined OAM beam is implemented via spatial filtering, and the radius of the truncated aperture $r$ is defined as $r = \gamma_{|m|}\lambda L / D$, where $D$ and $\gamma_{|m|}$ are the preliminarily determined total diameter and divergence coefficient. The optical field of the purified beam is expressed as

$$E_p(\rho,\psi,z=L) = E_c(\rho,\psi,z=L)circ(\rho/r). \tag{7}$$

Then, the normalized PIB for $r_{PIB} = r$ and the OAM spectrum of the spatially filtered beam can be obtained based on numerical calculation or experimental measurement. The normalized PIB of the combined OAM beam for $r_{PIB} = r$ can be represented in the form

$$PIB\Big|_{r_{PIB}=r} = \frac{\int_0^{2\pi}\int_0^{\infty}|E_p(\rho,\psi,z=L)|^2 \rho d\rho d\psi}{\int_0^{2\pi}\int_0^{\infty}|E_c(\rho,\psi,z=L)|^2 \rho d\rho d\psi}, \tag{8}$$

and the purity of the OAM mode with the TC of $m$ can be expressed as [56]

$$P_m = \frac{p_m}{\sum_{l=-\infty}^{+\infty} p_l}, \tag{9}$$

where

$$\begin{cases} p_l = \int_0^{+\infty}|a_l(\rho)|^2 \rho d\rho \\ a_l(\rho) = \frac{1}{\sqrt{2\pi}}\int_0^{2\pi} E_p(\rho,\psi,z=L)\exp(-il\psi)d\psi \end{cases}. \tag{10}$$

To evaluate the performance of the laser array system for structuring OAM beams, we define the generalized beam propagation factor (GBPF) as the performance metric, namely,

$$BPF^{(m)}\Big|_{\varepsilon_{PIB},\varepsilon_{Purity}} = \left(\frac{PIB\big|_{r_{PIB}=r}}{\eta_{|m|}}\right)^{\varepsilon_{PIB}} \cdot P_m^{\varepsilon_{Purity}}, \tag{11}$$

where $\varepsilon_{PIB}$ and $\varepsilon_{Purity}$ are the exponential indices corresponding to the PIB and OAM mode purity, respectively. The values of $\varepsilon_{PIB}$ and $\varepsilon_{Purity}$ depend on the specific scenario for the application of the laser array system. In our previous investigation, the indices ($\varepsilon_{PIB}$, $\varepsilon_{Purity}$) are chosen as (1, 1) for convenience, and the GBPF metric reduces to the higher-order BPF, of which the utility in optimizing the system design of on-off switching laser array based OAM emitters has been demonstrated [27]. For the energy transmission applications, the energy concentration of the combined beam is of dominant interest, and the OAM mode purity could be omitted, thus the indices ($\varepsilon_{PIB}$, $\varepsilon_{Purity}$) are set as (1, 0). The desirable TC of the laser array is zero, indicating that the coefficients of the zeroth-order (fundamental) Airy pattern are $\gamma_0 = 1.22$ and $\eta_0 = 83.78\%$, respectively. Therefore, the mathematical form of the GBPF metric reduces to

$$BPF^{(0)}\Big|_{1,0} = 1.19\, PIB\Big|_{r_{PIB}=1.22\lambda L/D}, \tag{12}$$

as the celebrated BPF metric that is widely used to characterize the beam quality of conventional CBC systems [54,55]. With the defined indices ($\varepsilon_{PIB}$, $\varepsilon_{Purity}$) and the normalized PIB and the OAM mode purity we have acquired in the fourth step, the performance metric of the system, namely GBPF, can finally be calculated.

Up to this point, we have presented the definition of GPBF metric and illustrated the general method to acquire the performance metric. It is worth mentioning that the acquisition of GBPF metric can be realized both theoretically and experimentally, which depends on the ways for the implementation of the third and fourth steps in the block diagram.

Specifically, when the performance metric is used for the design, optimization, and perturbative tolerance analysis of the laser array system prior to its construction, the third and fourth steps can be performed by using numerical calculations in a simulated environment. When the metric requires to be measured in experiments, here we provide a feasible schematic setup for the practical implementation of the GBPF metric measurement, as shown in Fig. 6.

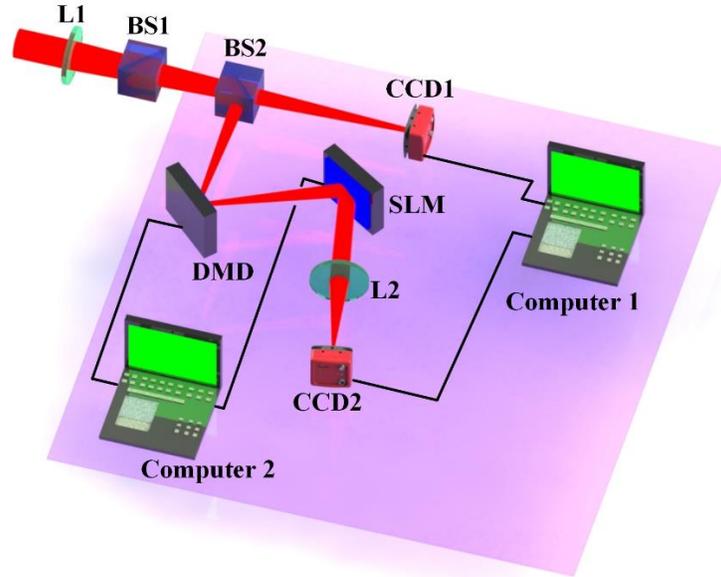

Fig. 6. Schematic setup of performance evaluation platform. (DMD: digital micromirror device; SLM: spatial light modulator; FL: focus lens; BS: beam splitter.)

The focused laser beams that transmit the BS1, as presented in Fig. 1, are delivered to the performance evaluation platform as the input. The focused beams are sampled by a beam splitter (BS2) into two parts. The transmitted part is directly focused to the CCD camera (CCD1) positioned at the Fourier plane of the focal lens L1 (see Fig. 1), and the detected intensity distribution of the combined OAM beam is sent to the computer (Computer 1) for the normalized PIB calculation. The reflected part successively propagates through a digital micromirror device (DMD) and SLM, which are driven by another computer (Computer 2). The DMD is capable of programmable intensity modulation and performs as the spatial filter to implement the beam purification, whereas the SLM, loaded with the Dammann optical vortex grating phase mask [22], is then used to simultaneously demultiplex the OAM modes. The modulated beam is focused by the focal lens L2 and the far-field diffraction pattern can be observed by the CCD camera (CCD2) placed at the focal plane of L2. The intensity signals collected by the CCD2 is sent to the Computer 1 for calculating the OAM spectrum. According to the calculated results of normalized PIB and OAM spectrum, the platform can realize the measurement of GBPF metric to characterize the performance of the combined OAM beam.

To summarize, we have introduced the model of the coherent fiber laser array system for structuring OAM beams in terms of the system architecture, mathematical representation of laser array, difficulties of using conventional performance metrics, and the definition and general acquisition method for the GBPF metric. In the following sections, the performance of the laser array system for structuring OAM beams is analyzed based on the GBPF metric and system model we have proposed, offering basic considerations for the system design and tolerance on perturbative factors.

### 3. Effect Analysis of Laser Sources Characteristics

We have introduced the performance metric and evaluation method of the coherent fiber laser array system for structuring OAM beams, ensuring the characterization of the system performance. According to the above-mentioned architectural model, the laser array system is composed of high-power laser sources, emitting array configuration, and dynamic control system as the three main sections. The characteristics of the three main sections have effect on the performance of the system, which deserves clear identification and analysis. In this section, we firstly study the effect of the laser sources characteristics on the system performance.

To begin with, the specific model of the coherent laser array we now investigate in the simulated environment is introduced. Without loss of generality, a 30-element, circular arranged coherent laser array is taken as an example. According to the general method to acquire the performance metric presented in the above section, we first determine the total diameter and desirable TC of the combined OAM beam. The intensity profile of the laser array at the source plane is displayed in Fig. 7(a), and the total diameter $D$ of the entire aperture for the laser array is given by 173 mm. To tailor OAM+2 and OAM+4 beams, the phase distributions at the source plane are shown in Figs. 7(b) and 7(c), respectively. With the knowledge of total diameter and desirable TCs (+2 and +4), the divergence coefficient and energy proportion for each case can be obtained by referring to the concept of higher-order Airy patterns, and for the cases of structuring OAM+2 and OAM+4 beams, the radii of the truncated aperture for beam purification are accordingly determined as 0.295 mm and 0.413 mm, respectively.

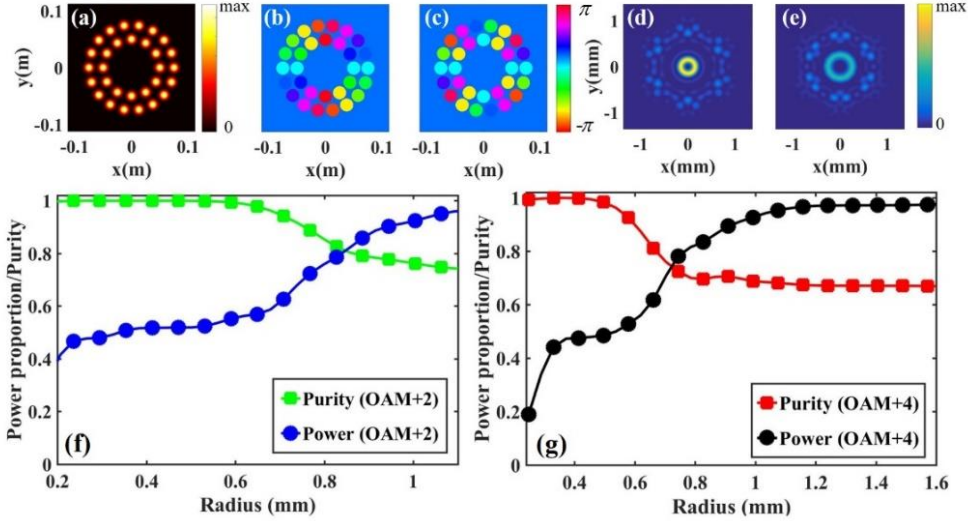

Fig. 7. Optical field characteristics of 30-element coherent laser array to tailor OAM+2 and OAM+4 beams. (a) Intensity profile of the laser array at the source plane. (b) and (c) exhibit the phase distributions at the source plane to tailor OAM+2 and OAM+4 beams, respectively. (d) and (e) are the intensity distributions of the combined OAM+2 and OAM+4 beams at the receiver plane, respectively. (f) and (g) depict the purity of the desirable OAM mode and power proportion as a function of the radius at the receiver plane, which correspond to the cases of structuring OAM+2 and OAM+4 beams, respectively.

In an ideal case, namely without system perturbations and misalignments, the intensity distributions of coherently combined OAM+2 and OAM+4 beams formed at the receiver plane are simulated, as shown in Figs. 7(d) and 7(e), respectively. Figure 7(f) exhibits the purity of the OAM+2 mode and power proportion as a function of the radius at the receiver plane, which corresponds to the customization of OAM+2 beam. For the case of structuring OAM+4 beam, the OAM mode purity and power proportion as a function of the radius at the receiver plane are depicted in Fig. 7(g). Utilizing the circular apertures with the radii of 0.295 mm and 0.413 mm to purify the combined OAM beams, the mode purity and energy proportion of the purified OAM+2 beam are: in turn, calculated as 99.95% and 47.90%, and the mode purity and energy proportion that correspond to the purified OAM+4 beam are 99.79% and 47.45%, respectively. For a variety of applications for optical OAM beams, such as data transmission and rotating detection, OAM mode purity is of great significance, and it is generally preferred to generate the OAM beam of the mode purity as high as possible to ensure its utility. In this work, the indices ($\varepsilon_{PIB}$, $\varepsilon_{Purity}$) of the performance metric GBPF are thus set as (0, 1). Therefore, in terms of the definition given by Eq. (11), the GBPF of the ideal coherently combined OAM+2 beam tailored with the 30-element laser array is 0.9995, and for the ideal combined OAM+4 beam, the value of the GBPF metric is calculated as 0.9979.

Here we have determined the value of GBPF metric for the situation of ideal coherent combination, indicating that the influence of system perturbations and misalignments can be quantitatively characterized by analyzing the loss of GBPF. For the characteristics of laser sources, the power (amplitude) variations and depolarization among the laser array channels are the two key factors that could result in the performance degradation of the combined OAM beams. The effect analysis and results for the nonuniformities of channel powers and polarizations are presented as follows.

### 3.1 Effect of power variations

Power nonuniformity of laser sources is a common problem for the practical implementation of laser beams combination. Especially for coherent laser beam combination, it has been demonstrated that the power variations among array channels contribute to the loss of combining efficiency [59]. In fact, the nonuniformity of powers (amplitudes) not only affects the intensity distribution formed in the far-field, but also distorts the wavefront of the combined beam. Therefore, when the combined OAM beams with both exotic spatial intensity profiles and wavefronts are considered, investigating the effect of power variations is highly required.

To characterize the dependences of system performance on the power nonuniformity of laser sources, Monte-Carlo simulations have been conducted for both the cases of structuring OAM+2 and OAM+4 beams. We assume that the powers of all channels are $P$ for the ideal combination situation, and $\delta P/P$ denotes the normalized power variations. For different value of the root-mean-squared (RMS) power variations that ranges from 0 to 0.5, one hundred sets of powers are randomly generated, and the averaged GBPF metrics are calculated and recorded, as displayed in Fig. 8(a). One can observe that the GBPF metrics for both the cases of tailoring OAM+2 and OAM+4 beams decrease as the value of RMS power variations increases, and the descending trend of GBPF is more apparent for structuring the OAM beam with a larger absolute value of TC. Figures 8(b1)-8(b5) exhibit the averaged intensity distributions (calculated 100 times) of the combined OAM+2 beams when the RMS power variations equal to 0.1, 0.2, 0.3, 0.4, and 0.5, respectively, and the averaged intensity distributions of the combined OAM+4 beams are, in turn, shown in Figs. 8(c1)-8(c5). When the value of RMS power variations gradually increases to 0.5, slight intensity spreading from the main ring of the combined OAM beam to the inner vortex core and the outer sidelobes can be observed. Based on the simulated results, it can be summarized that the power variations affect the performance of combined OAM beams, and the performance degradation is more obvious for the higher-order OAM modes. To ensure the <1% loss of GBPF metric, the RMS power variations of the 30-element coherent laser array should not exceed 0.2 for structuring OAM+4 beams and 0.3 for structuring OAM+2 beams.

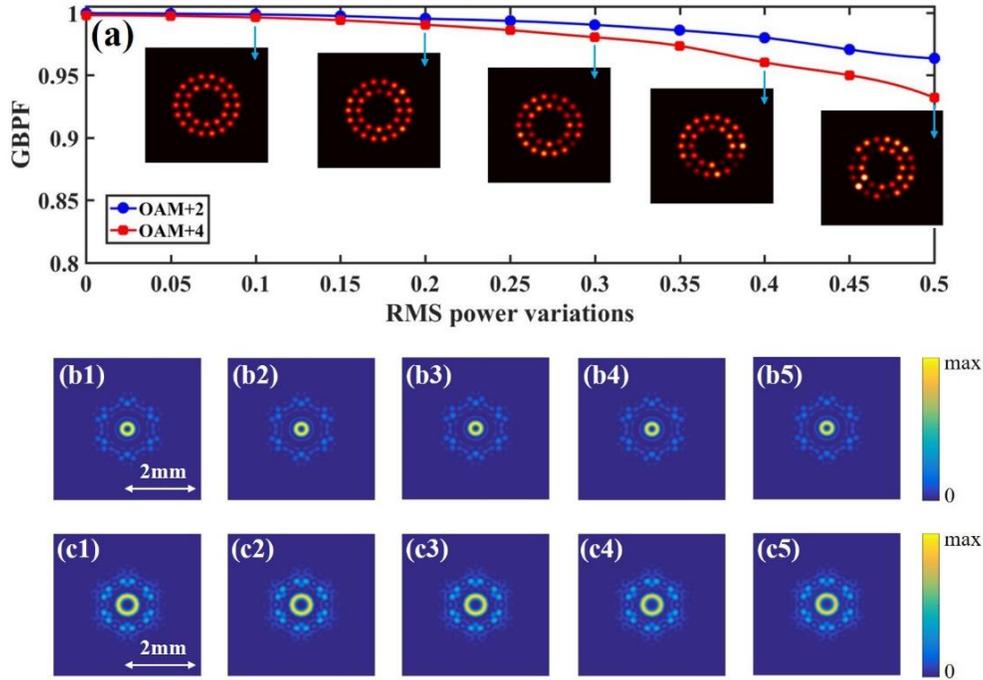

Fig. 8. Effect of power nonuniformity on performance of combined OAM beams. (a) GBPF metric as a function of RMS power variations for structuring OAM+2 and OAM+4 beams. The inset figures are the typical intensity distributions of laser arrays at the source plane when the RMS power variations are set as different values. Averaged intensity distributions (calculated 100 times) of the combined OAM+2 beams when the RMS power variations equal to (b1) 0.1, (b2) 0.2, (b3) 0.3, (b4) 0.4, and (b5) 0.5. (c1)-(c5) depict the averaged intensity distributions of the combined OAM+4 beams when the RMS power variations equal to 0.1, 0.2, 0.3, 0.4, and 0.5, respectively.

Moreover, the damage of array elements, as a special case of power variations, deserves analysis and discussion. We first illustrate whether the damage of the array element lying on the inner subarray or the outer subarray has a greater impact on the performance of the combined OAM beams. Figures 9(a1) and 9(a2) depict the intensity distributions of the laser arrays with a damaged element lying on the inner radial subarray and outer radial subarray at the source plane, respectively. When the element of the inner subarray is damaged, the far-field intensity distribution of the combined OAM+2 beam is shown in Fig. 9(b1), and the GBPF metric is calculated as 0.9868. The far-field intensity distribution of the combined OAM+4 beam is presented in Fig. 9(c1), and the GBPF metric is 0.9737. In comparison, for the case of the element lying on the outer subarray is damaged, the far-field intensity distributions of the combined OAM+2 and OAM+4 beams are displayed in Figs. 9(b2) and 9(c2), respectively. The GBPF metrics for the customization of OAM+2 and OAM+4 beams are calculated as 0.9863 and 0.9751, respectively. The results suggest that the laser array system is more sensitive to the damage of array element when the higher-order OAM beam is tailored, and the damage on the outer subarray leads to more serious performance degradation when compared to the damage on the inner subarray.

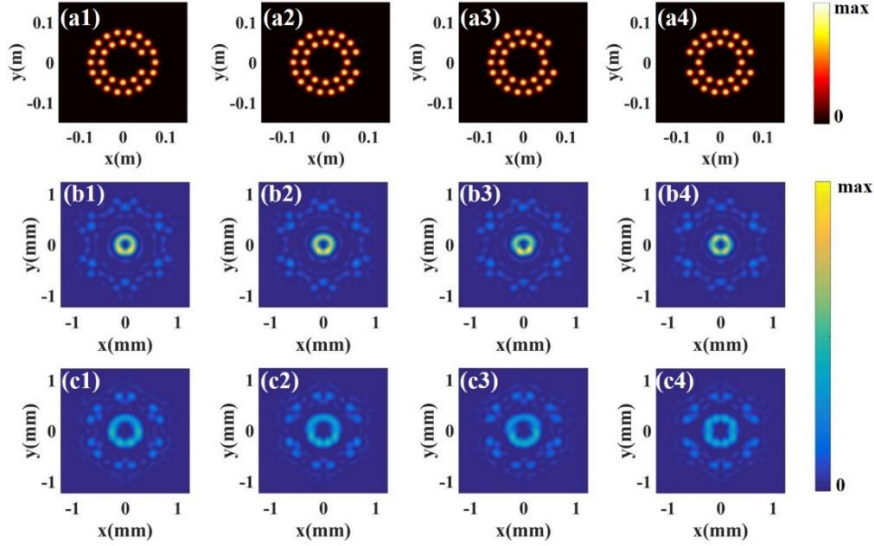

Fig. 9. Effect of damaged array elements on the performance of combined OAM beams. Near-field intensity distributions of the laser arrays with (a1) a damaged element lying on the inner radial subarray, (a2) a damaged element lying on the outer radial subarray, (a3) two adjacent damaged elements lying on the outer radial subarray, and (a4) two symmetric damaged elements lying on the outer radial subarray. (b1)-(b4) exhibit the far-field intensity distributions of the OAM+2 beams formed by the coherent combination of the laser arrays presented in (a1)-(a4), respectively. (c1)-(c4) are the far-field intensity distributions of the OAM+4 beams that correspond to (a1)-(a4), respectively.

Then, the damage of two array elements is analyzed as well. When the adjacent array elements lying on the outer radial subarray are damaged [see Fig. 9(a3)], the far-field intensity distributions of the combined OAM +2 beam with a GBPF metric of 0.9631 and the combined OAM+4 beam with a GBPF metric of 0.9485 are shown in Figs. 9(b3) and 9(c3), respectively. In contrast, when the damaged elements are positioned symmetrically [see Fig. 9(a4)], the GBPF metrics for the combined OAM+2 beam [see Fig. 9(b4)] and combined OAM+4 beam [see Fig. 9(c4)] are calculated as 0.9749 and 0.9543, respectively. Intensity distortions of the combined OAM beams can be observed, and the loss of GBPF metric becomes more obvious when two elements of the laser array are damaged. Besides, the case of damage in the adjacent elements would cause a more serious performance degradation.

In a nutshell, we have analyzed the effect of power variations on the performance of combined OAM beams. For the special case of power variations, namely the damage of array elements, more significant performance degradation would be induced. The results and findings could be useful for the optimization of high-power laser sources.

### 3.2 Effect of depolarization

In coherent fiber laser array systems to generate high-brightness combined beam (conventional CBC system) and to tailor scalar structured light beams, the laser beams of all channels are linearly polarized, and the polarization states are matched to ensure the efficient interference in the desirable polarization direction. However, in practical implementations, the presence of depolarization among the laser array channels tends to degrade the performance of the combined beam. For conventional CBC systems, the effect of polarization mismatch on the combining efficiency has been analyzed [54,60]. For the laser array systems that tailor OAM beams with more complex optical field structures, utilizing the proposed performance metric GBPF to characterize the effect of the laser sources depolarization on the combined OAM beams is also important but lacks exploration, which we now investigate.

Figure 10 depicts the dependences of the GBPF metrics on the polarization errors of the array channels for the customization of OAM+2 (blue curve) and OAM+4 beams (red curve). For each value of polarization errors, the GBPF metric is calculated as the averaged value for one hundred Monte-Carlo simulations. The results indicate that with the increase of the polarization errors of the array channels, the GBPF metrics are decreased for the two cases. Specifically, the GBPF drops from 0.9995 to 0.9966 for the case of tailoring OAM+2 beams, and drops from 0.9979 to 0.9923 for structuring OAM+4 beams, when the polarization errors increase from zero to 20 degree. The inset figures exhibit the averaged intensity distributions of the combined OAM+2 beam and OAM+4 beam when the polarization errors equal to 20 degree, which are similar to the results for ideal coherent combination shown in Figs. 7(d) and 7(e). Hence, we can conclude that the tolerance of the laser sources depolarization on the performance of combined beams is relatively high for the laser array systems, and the performance degradation for the combined OAM beam with a larger absolute value of TC is slightly more obvious.

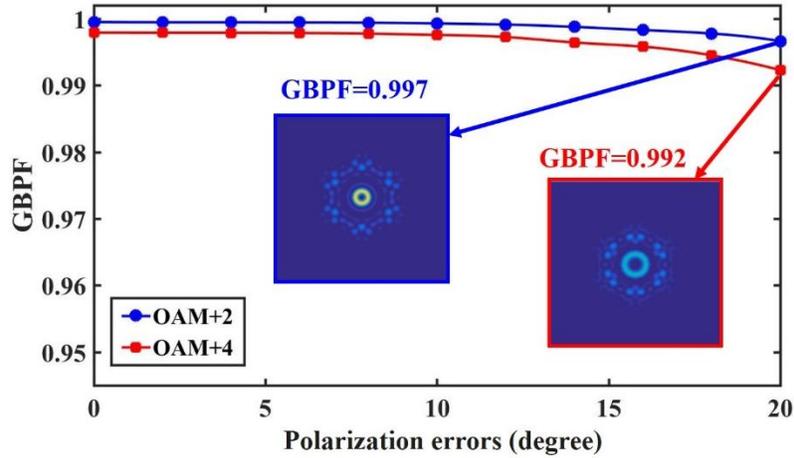

Fig. 10. GBPF metrics versus the polarization errors of the array channels for the 30-element coherent laser array systems to tailor OAM+2 beam (blue) and OAM+4 beam (red). The inset figures are the averaged far-field intensity distributions of the combined OAM beams when the polarization errors equal to 20°.

## 4. Effect Analysis of Emitting Array Configuration

In coherent fiber laser array systems, the emitting array configuration is an important section that determines the intensity distribution of the optical field at the source plane and further determines the intensity and phase distributions of the combined beam in the far-field. Accordingly, the design of emitting array configuration has effect on the performance of the combined OAM beam, which we now investigate. In this section, the effect of emitting array configuration characteristics on the performance of combined beam is analyzed in terms of three main aspects, namely the geometric arrangement of the laser array, filling of the conformal aperture, and the filling of subaperture.

### 4.1 Geometric arrangement

In typical coherent fiber laser array systems for brightness enhancement and spatial light structuring, the beamlets are always arranged by the emitting array configuration in a centrally symmetric shape. The symmetrical geometric arrangement of beamlets mainly includes two categories, namely the hexagonal arrangement and circular arrangement. In this work, we focus on the scenario of tailoring OAM beams with coherent fiber laser array systems, and the laser array of circular arrangement is considered and analyzed. It is worth mentioning that for the conventional CBC systems, namely forming the energy concentrated beam pattern in the far-field by coherently combining multiple beamlets of the laser array, close-packed hexagonal arrangement is widely employed as well [61-63]. In fact, the hexagonal arrangement has been also utilized for structuring OAM beams in femtosecond regime. In the 61-channel, hexagonal arranged coherent combining system, the generation of OAM beams with the TC ranges from 1 to 3 has been demonstrated [34]. Indeed, compared to the circular arrangement, the close-packed hexagonal arrangement is more compact thus is beneficial for sidelobes suppression. In other words, the hexagonal arrangement suggests a more concentrated energy of the combined beam and a higher combining efficiency, and the intuitive idea for designing the geometric arrangement of the emitting laser array for spatial light structuring is to directly use the hexagonal arrangement. However, for the scenario of tailoring OAM beams, the mode purity is vital and should not be neglected. Hence, it is necessary for system design to revisit and discuss the choice of emitting laser array arrangement between the hexagonal and circular types.

Figure 11 exhibits the optical fields of the emitting laser arrays in terms of the hexagonal and circular arrangements and the corresponding combined OAM beams. For the emitting laser array of close-packed hexagonal arrangement and consists of 30 elements, the intensity distribution at the source plane is shown in Fig. 11(a). To generate the OAM beams with the TCs of 1, 2, 3, and 4, the phase distributions of the laser array at the source plane are presented in Figs. 11(c1)-11(c4), respectively, and the intensity distributions of the combined OAM+1, OAM+2, OAM+3, and OAM+4 beams in the far-field are shown in Figs. 11(d1)-11(d4), respectively. In comparison, Fig. 11(b) depicts the intensity distribution of the circular arranged laser array at the source plane, which also contains 30 beamlets. For the case of circular arrangement, the phase distributions of the laser array at the source plane to generate OAM+1, OAM+2, OAM+3, and OAM+4 beams are displayed in Figs. 11(e1)-11(e4), respectively. Correspondingly, the far-field intensity distributions of the combined OAM beams with the TCs of 1, 2, 3, and 4 are shown in Figs. 11(f1)-(f4), respectively. Compared to the circular arrangement case, the laser array of hexagonal arrangement is more compact, and the spacing between the adjacent beamlets is uniform. In the far-field, the combined OAM beam has a hexagonal-shaped and dark-hollow central pattern and surrounding sidelobes. Along the azimuthal direction, it can be apparently observed that the intensity distribution of the central pattern has the non-uniform characteristic. In contrast, as for the combined OAM beam of the circular arranged laser array, a main ring with a uniform intensity distribution along the azimuthal direction is always formed when the TC varies from 1 to 4.

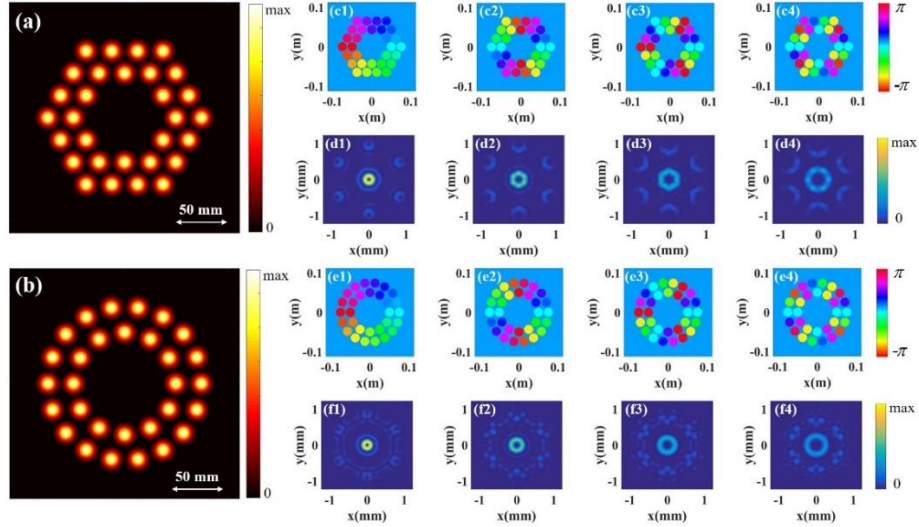

Fig. 11. Optical fields of emitting laser arrays and combined OAM beams. Intensity distributions of the (a) 30-element hexagonal arranged laser array and (b) 30-element circular arranged laser array at the source plane. For the hexagonal arrangement, (c1), (c2), (c3), and (c4) are the phase distributions of the laser array at the source plane to generate the OAM+1, OAM+2, OAM+3, and OAM+4 beams, respectively. (d1), (d2), (d3), and (d4) present the corresponding far-field intensity distributions of the combined OAM+1, OAM+2, OAM+3, and OAM+4 beams, respectively. For the circular arrangement, the phase distributions of the laser array at the source plane to generate the OAM+1, OAM+2, OAM+3, and OAM+4 beams are shown in (e1), (e2), (e3), and (e4), respectively. (f1), (f2), (f3), and (f4) are the far-field intensity distributions of the combined OAM+1, OAM+2, OAM+3, and OAM+4 beams, respectively.

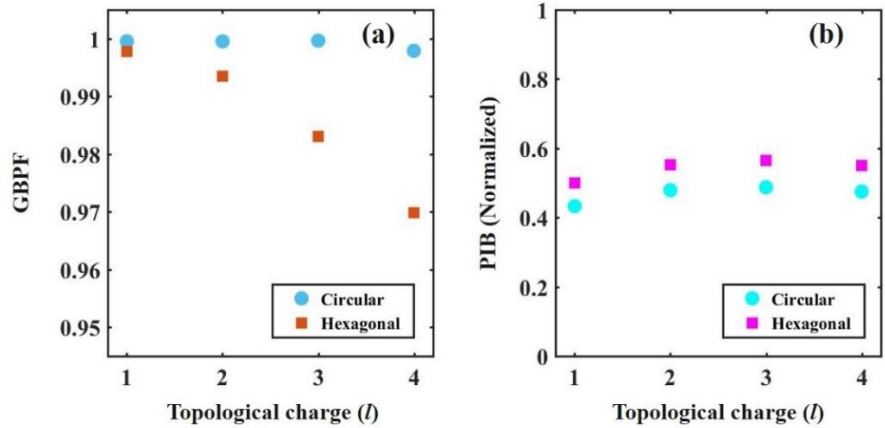

Fig. 12. Comparison between circular and hexagonal arrangements in terms of GBPF metric and PIB. (a) GBPF metric as a function of the TC for the OAM beam tailored by the circular and hexagonal arranged laser arrays. (b) Normalized PIB as a function of the TC for the OAM beam tailored by the circular and hexagonal arranged laser arrays.

To further quantitatively compare the effect of circular and hexagonal arrangements on the combined OAM beam, the GBPF metric to evaluate the performance of the combined beam has been calculated for the two cases when the TC of the combined OAM beam ranges from 1 to 4, and the results are presented in Fig. 12(a). For the case of circular arrangement, the calculated GBPF always maintains at relative high values (>0.99), and is slight decreased from 0.9995 to 0.9979 when the TC varies from 1 to 4. However, when the beamlets are arranged in hexagonal shape at the source plane, the GBPF metrics are calculated as 0.9978, 0.9935, 0.9830, and 0.9699, which correspond to the performance of combined OAM+1, OAM+2, OAM+3, and OAM+4 beams, respectively. The results indicate that the hexagonal arrangement not only cause the non-uniform characteristic of the central pattern of the combined OAM beam, but also affect the OAM mode purity. With the increase of

TC, the advantage of circular arrangement in terms of GBPF and mode purity becomes more apparent. Besides, the energy concentration of the combined OAM beams for the circular and hexagonal arrangements is evaluated by calculating the normalized PIB. For the two geometric arrangements, the circumscribe circles are the same, and the total diameters of the entire aperture are both 173 mm. Here, we can determine the value of normalized PIB based on the coefficients of higher-order Airy pattern, total diameter, and the far-field intensity distribution of the combined OAM beam, and the results of the calculated PIB for the cases of circular and hexagonal arrangements are shown in Fig. 12(b). The normalized PIB for the hexagonal arrangement is always relatively higher than the value for the circular arrangement, indicating that the hexagonal arrangement offers a more concentrated energy distribution, which is consistent with the conventional insights.

To conclude, the comparison between the circular arrangement and hexagonal arrangement has been carried out. On the one hand, the circular arrangement of the emitting laser array can ensure the uniform characteristic of the main ring of combined OAM beam along the azimuthal direction. One the other hand, when compared to the hexagonal arrangement, circular arrangement can improve the purity of the desirable OAM mode, and the GBPF as the metric to evaluate the performance of combined beam would not decrease with the increase of TC. Accordingly, even though the close-packed hexagonal arrangement is beneficial to enhance the energy concentration of the combined beam, as for the scenarios that require the OAM beam of high purity, circular arrangement is the preferred better choice.

### 4.2 Filling of conformal aperture

In emitting laser array configurations, the filling of conformal aperture is an important factor that associates with the far-field intensity patterns. In the view of conventional coherent combining for brightness enhancement, the filling of conformal aperture describes the compactness of the laser array and influences the energy proportion for the mainlobe of the combined beam [30,53,54]. Recently, it has been demonstrated that modulating the filling of conformal aperture is capable of spatial beam shaping and can especially customize the structure of optical vortex lattices [45]. Here, we focus on the coherent fiber laser array system for tailoring OAM beams and analyze the effect of conformal aperture filling on the system performance, of which the results could provide useful references on the system design.

To characterize the filling of conformal aperture, the concept of fill factor has been widely used, which has various forms of definitions such as the conformal fill factor [53], lenslet fill factor [64], and areal fill factor [30]. As for the hexagonal arranged laser array, the conformal fill factor is defined as $d_n / d$, where $d$ accounts for the diameter of the circular subaperture, and $d_n$ represents the distance between the centers of neighboring subapertures. However, as for the circular arranged laser array, the structure is not periodic, and the beamlets lying on the outer radial subarray are closer when compared to the beamlets of the inner radial subarray [28]. Along the radial direction, the distances of the adjacent beamlets always equal to the structure parameter $R$, which has been introduced in the section 2.2. Hence, we are motivated to modified the conformal fill factor to the radial fill factor with the mathematical form of $R / d$, which is more appropriate to characterize the compactness of the circular arranged laser array.

Figures 13(a) and 13(b) exhibit the performance metric GBPF of the combined OAM beams as a function of the radial fill factor $R / d$, which correspond to the cases of TC = +2 and TC = +4, respectively. The intensity distributions of the laser arrays at the source plane with different radial fill factors are displayed in the inset figures. As the radial fill factor increases, the arrangement of beamlets gradually evolves from compact to sparse, whereas the GBPF metric always exceeds 0.993 for both the cases of structuring OAM+2 and OAM+4 beams. The reason for the almost invariant property of the performance metric can be illustrated as the indices ($\varepsilon_{PIB}$, $\varepsilon_{Purity}$) of the GBPF metric are defined as (0, 1), since the OAM mode purity is of dominant interest. Although the distance between the neighboring beamlets of the laser array varies, high mode purity of the combined OAM beam after beam purification (implemented by spatial filtering, see section 2.4) is ensured throughout the variance of the combined optical field. Figures 13(c1)-13(c6) depict the intensity distributions of the combined OAM+2 beams after beam purification, which correspond to the cases of $R / d$ = 1, 1.2, 1.4, 1.6, 1.8, and 2, respectively. The intensity distributions of the combined OAM+4 beams after beam purification when the radial fill factor $R / d$ equals to 1, 1.2, 1.4, 1.6, 1.8, and 2, in turn, are shown in Figs. 13(d1)-13(d6). It can be observed that for a determined TC of the combined OAM beam, doughnut-shaped intensity patterns can be formed after beam purification, and the intensity peak almost remains the same as the filling of the conformal laser array varies. The less compact laser array has a larger total diameter of the entire conformal aperture, and accordingly, the size of the purified OAM beam is smaller.

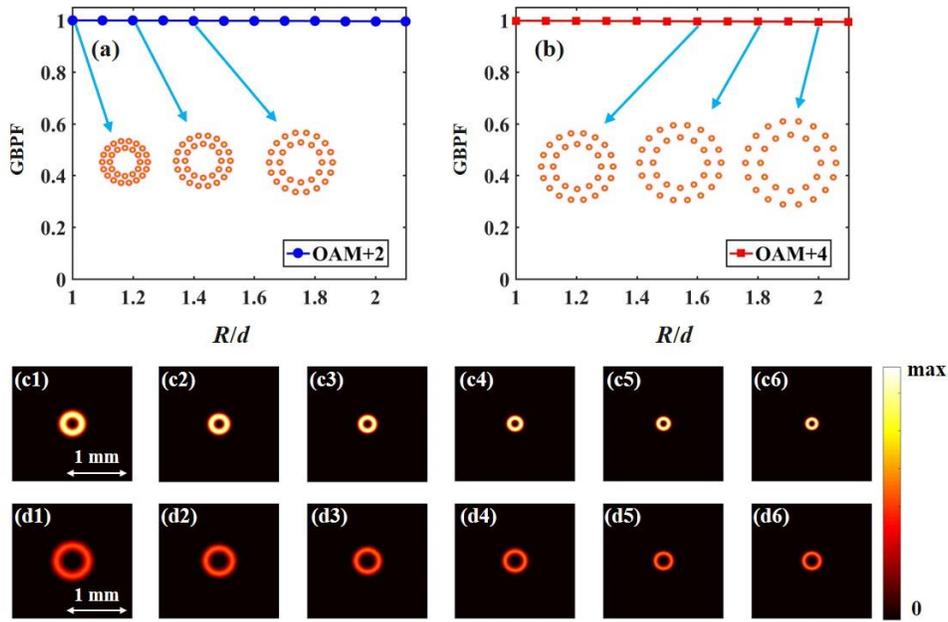

Fig. 13. Effect of conformal aperture filling on performance of combined OAM beams. (a) GBPF metric as a function of radial fill factor $R/d$ for structuring OAM+2 beams. (b) GBPF metric as a function of radial fill factor $R/d$ for structuring OAM+4 beams. The inset figures are the intensity distributions of laser arrays with different radial fill factors. (c1)-(c6) depict the intensity distributions of the combined OAM+2 beams after beam purification, which correspond to the cases of $R/d = 1$, $R/d = 1.2$, $R/d = 1.4$, $R/d = 1.6$, $R/d = 1.8$, and $R/d = 2$, respectively. (d1)-(d6) are, in turn, the intensity distributions of the combined OAM+4 beams after beam purification when $R/d$ equals to 1, 1.2, 1.4, 1.6, 1.8, and 2.

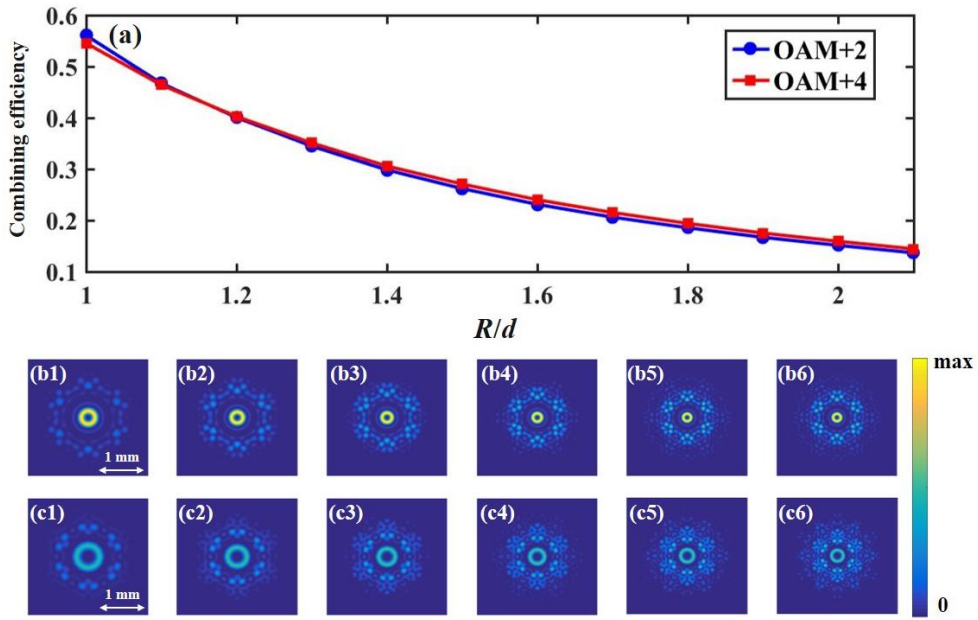

Fig. 14. (a) Evolution curves of combining efficiency when the radial fill factor $R/d$ varies for structuring OAM+2 (blue) and OAM+4 beams (red). For the laser arrays with the radial fill factor of $R/d = 1$, $R/d = 1.2$, $R/d = 1.4$, $R/d = 1.6$, $R/d = 1.8$, and $R/d = 2$, the far-field intensity distributions of the combined OAM+2 beams (without beam purification) are displayed in (b1)-(b6), and the far-field intensity distributions of the combined OAM+4 beams (without beam purification) are presented in (c1)-(c6), respectively.

According to the above results of GBPF metric, we can conclude that with the beam purification, the laser arrays of different radial fill factors are all capable of structuring high performance OAM beams. Despite the high OAM mode purity of the combined beam with purification, sidelobes of the combined beam that are spatially filtered out, which inevitably causes the energy loss of the system. Different filling of conformal aperture leads to the diverse intensity distributions of the combined OAM beams, thus the energy proportions of the sidelobes are different. Here we define the combining efficiency as the percentage of energy in the purified beam, and we confine the OAM mode purity of the purified beam to above 99%. The defined combining efficiency can characterize the energy loss for the process of OAM beams tailoring. Actually, the combining efficiency we defined is linearly proportional to the GBPF metric with the indices ($\varepsilon_{PIB}$, $\varepsilon_{Purity}$) of (1, 0). Figure 14(a) exhibits the combining efficiency as a function of radial fill factor for the customization of OAM+2 (blue) and OAM+4 beams (red). The far-field intensity distributions of the combined OAM+2 beams (without beam purification) are shown in Figs. 14(b1)-14(b6), which correspond to the laser arrays with the radial fill factors of 1, 1.2, 1.4, 1.6, 1.8, and 2, respectively. In Figs. 14(c1)-14(c6), from left to right, they are the far-field intensity distributions of the combined OAM+4 beams (without beam purification) for the cases of $R/d$ = 1, 1.2, 1.4, 1.6, 1.8, and 2, respectively. When the distance between the adjacent beamlets is enlarged, the sidelobes of the combined OAM beams become more apparent, and therefore, the decrease of combining efficiency can be observed. The decreased trend of combining efficiency for structuring OAM+2 beams is approximately coincided with the case of structuring OAM+4 beams. Based on our findings, we can summarize that the laser array of more compact arrangement (lower radial fill factor) is more conducive to the suppression of sidelobes and can alleviate the energy loss of beam purification, indicating a higher combining efficiency could be reached. Hence, to design the emitting laser array configuration, comprehensively driven by the preservation of GBPF metric and enhancement of combining efficiency, the beamlets arrangement of high compactness is preferred, and the radial fill factor should be engineered to approach 1.

### 4.3 Filling of subaperture

Analogous to the filling of conformal aperture, the filling of subaperture is another important consideration which the design of laser array system depends. In previous studies of conventional CBC systems, several definitions have been proposed and used to characterize the filling of subaperture, which include the subaperture fill factor [53] and Gaussian fill factor [64,65]. For the benefit of convenience and utility, we employ the subaperture fill factor to analyze the effect of subaperture filling on the performance of the coherent fiber laser array system for tailoring OAM beams, and the subaperture fill factor is defined as $2w_0/d$, where $w_0$ and $d$ account for the waist width of each Gaussian beamlet and the diameter of the circular subaperture.

We first analyze whether the filling of subaperture would affect the generation of high purity OAM beams. The performance metric GBPF, for the cases of structuring OAM+2 and OAM+4 beams, as a function of the subaperture fill factor is depicted in Figs. 15(a) and 15(b), respectively. When the subaperture fill factor varies, the GBPF metric of the combined OAM beams approximately maintains the same value. For structuring the OAM beams with the TCs of +2 and +4, the GBPF can always exceed 0.999 and 0.997, respectively. The results reveal the fact that when we utilize the GBPF with the indices ($\varepsilon_{PIB}$, $\varepsilon_{Purity}$) of (0, 1) as the performance metric to characterize the laser array system, the OAM mode purity is of dominant interest, and the variation of subaperture filling almost has no impact on the performance of the combined OAM beam with beam purification. In other words, high OAM mode purity of the combined beam (after beam purification) is ensured throughout the change of subaperture fill factor. Figures 15(c1)-15(c4) exhibit the intensity profiles of the laser arrays at the source plane with the subaperture fill factors of typical values, namely $2w_0/d = 0.5$, $2w_0/d = 1$, $2w_0/d = 1.5$, and $2w_0/d = 2$, respectively. Correspondingly, the combined OAM+2 beams and OAM+4 beams after beam purification are shown in Figs. 15(d1)-15(d4), and Figs. 15(e1)-15(e4), respectively. One can see that the laser array with a lower subaperture fill factor is equivalent to the laser array with a higher radial fill factor, as we have discussed in the above section. With the increase of subaperture fill factor, the intensity distribution of each subaperture gradually evolves from the Gaussian pattern to the truncated quasi-plane wave illumination. Despite the change of intensity distributions at the source plane, doughnut-shaped intensity patterns can always be formed after beam purification for the generation of OAM+2 and OAM+4 beams, and the beam size almost remains the same value for each case as well.

Indeed, the laser array with a higher subaperture fill factor is closer to the annular aperture truncated plane wave illumination. In terms of the definition for higher-order Airy patterns, we can illustrate that through continuously enlarging the subaperture fill factor, the laser array at the source plane would gradually approach the optical field that forms the higher-order Airy patterns in the far-field, indicating a more significant efficiency for sidelobes suppression. However, during the process of enlarging the subaperture fill factor, more energy loss of the laser sources would be caused by the truncation of circular subapertures. Since the filling of subaperture would not affect the OAM mode purity of the combined beam with beam purification, we now elucidate the effect of subaperture fill factor on the total efficiency of the fiber laser array system. The total efficiency $\eta_{total}$ is defined as $\eta_{total} = (1 - L_{sub})(1 - L_{filter})$, where $L_{sub}$ and $L_{filter}$ denote the percentages of energy loss owing to the subaperture truncation at the source plane and the spatial filtering for beam purification. The efficiency of beam purification, namely $(1 - L_{filter})$, is linearly proportional to the GBPF metric with the indices ($\varepsilon_{PIB}$, $\varepsilon_{Purity}$) of (1, 0). According to the above analysis, there is a tradeoff for alleviating the energy loss caused by subaperture truncation and beam purification, thus we are inspired to optimize the subaperture fill factor.

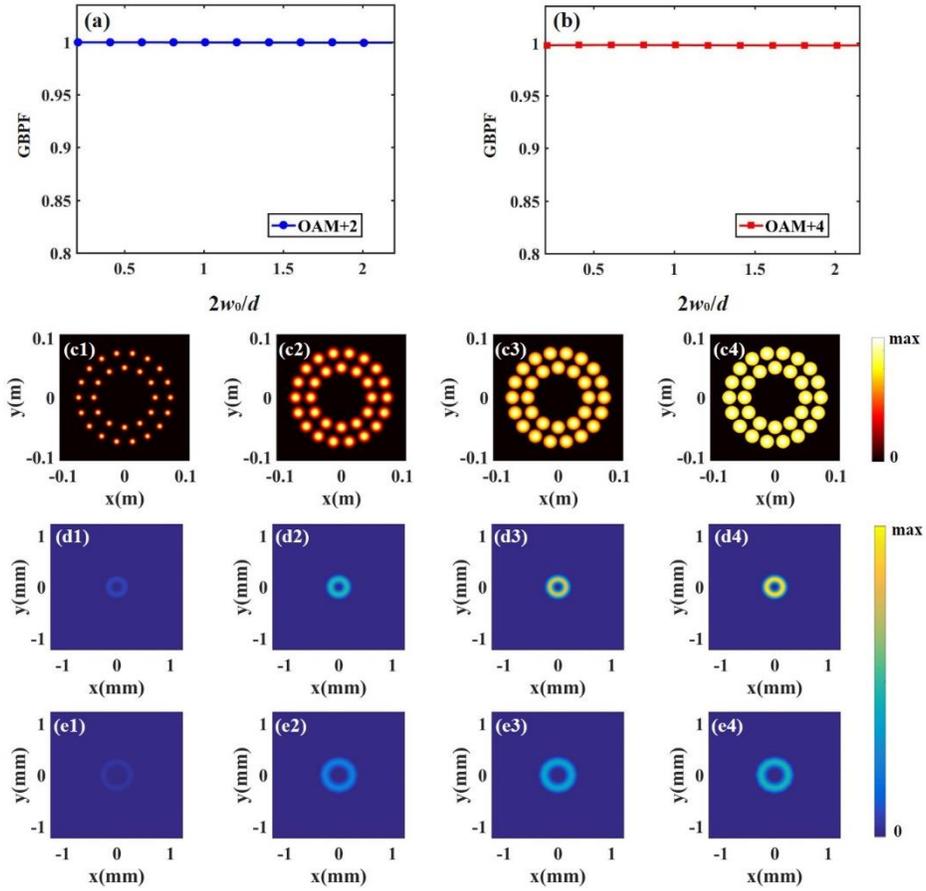

Fig. 15. Effect of subaperture filling on performance of combined OAM beams. (a) GBPF metric as a function of subaperture fill factor $2w_0/d$ for structuring OAM+2 beams. (b) GBPF metric as a function of subaperture fill factor $2w_0/d$ for structuring OAM+4 beams. Intensity distributions of the laser arrays at the source plane with the subaperture fill factor of (c1) $2w_0/d = 0.5$, (c2) $2w_0/d = 1$, (c3) $2w_0/d = 1.5$, and (c4) $2w_0/d = 2$. (d1), (d2), (d3), and (d4) denote the intensity distributions of the combined OAM+2 beams after beam purification when the subaperture fill factor equals to 0.5, 1, 1.5, and 2, respectively. (e1), (e2), (e3), and (e4) are the intensity distributions of the combined OAM+4 beams after beam purification, which correspond to the cases of $2w_0/d = 0.5$, $2w_0/d = 1$, $2w_0/d = 1.5$, and $2w_0/d = 2$, respectively.

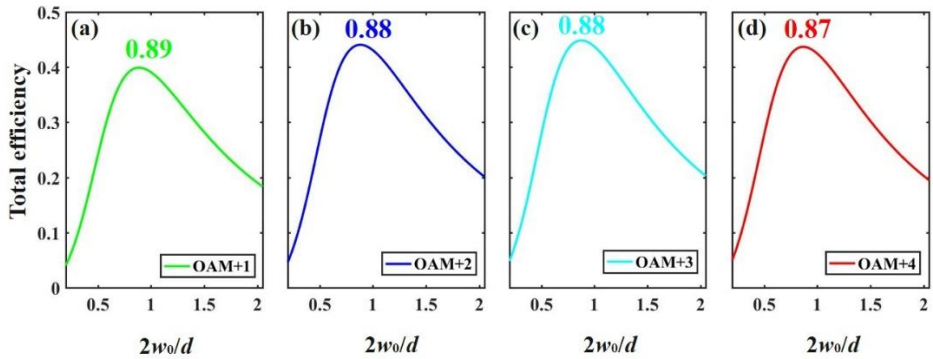

Fig. 16. Optimization of subaperture fill factor for system design. Total efficiency as a function of subaperture fill factor $2w_0/d$ for structuring (a) OAM+1, (b) OAM+2, (c) OAM+3, and (d) OAM+4 beams.

The optimization of subaperture fill factor for system design has been carried out, and the results are presented in Fig. 16. The total efficiency as a function of subaperture fill factor is depicted in Figs. 16(a)-16(d), which corresponds to the generation of OAM+1, OAM+2, OAM+3, and OAM+4 beams, respectively. For the four cases, the common phenomenon could be found that the total efficiency initially increases as the subaperture fill factor increases, and after the maximum value of total efficiency is reached, the total efficiency gradually drops with the further increase of the subaperture fill factor. Similar trend has been observed in the subaperture optimization for conventional CBC systems [53,65]. Our results further indicate that the optimized values of subaperture fill factors depend on the desirable TC of the combined OAM beam, and the conventional coherent combining corresponds to the special case of TC= 0. In this work, to tailor the OAM beams with the TCs of +1, +2, +3, and +4, the subaperture fill factors are optimized as 0.89, 0.88, 0.88, and 0.87, respectively.

## 5. Effect Analysis of Dynamic Control System

In the above sections, the characteristics of laser sources and emitting laser array configuration and their effect on the system performance have been investigated and discussed. Once the laser sources and emitting laser array configuration are designed, the intensity distributions of the laser array at the source plane could be determined. As for conventional CBC systems, the wavefronts of array elements are matched to fit a whole plane wave, thus ensuring the constructive interference in the far-field [29]. As for the laser array systems that tailor structured light beams, even though the intensity distribution at the source plane is determined, the controllable wavefronts of array elements enable the laser arrays to fit the exotic beams with various spatial structures at the source plane.

The programmable customization of the array element wavefronts relies on the dynamic control system, which mainly possesses the capacity of piston phase and tip-tilt modulations. For one thing, the dynamic control system can implement active wavefront shifting of each array element to realize the flexible spatial mode switching of the combined beams. For another thing, the aberrations of piston and tilt, as the first two terms of the Zernike polynomials, are the main concerns of the laser array system that affect the performance of the combined beams. Hence, the piston and tilt wavefront distortions are highly required to be compensated in real time by operating the dynamic control system. In the following sections, the effect of dynamic control system in terms of piston phase and tip-tilt wavefront control on the system performance is analyzed.

### 5.1 Piston phase control

In coherent laser array systems for structuring OAM beams, the piston phases of array elements are controlled to compensate the dynamic phase errors and realize the desirable phase shifting. Specifically, when the system is operated at high-power level, the thermal and environmental fluctuations induced piston phase errors always exist and seriously degrades the performance of combined beams, and therefore, the piston phase errors should be compensated via operating the dynamic control system in closed loops [60,66]. Besides, to fit the helical phase structure with laser array elements at the source plane for tailoring OAM beams in the far-field, active piston phase shifting of each channel is also required. To realize the compensation of piston phase errors and implementation of active phase shifting at the same time, diverse methods have been developed and demonstrated for the dynamic piston phase control of laser array systems, which include the interferometric [34,42] or heterodyne measurement techniques [43], machine learning methods [36,67], and the conventional optimization algorithm based methods with the assistance of OAM mode sorting [68].

Despite the advances in piston phase control methods, the dynamic control system that applies control signals to the array elements cannot completely compensate the phase errors, and therefore, residual phase errors are still exist during the closed-loop phase control process. The residual phase errors result in the deviation of the combined beam of the practical laser array system and the ideal combined OAM beam of theoretical prediction, which could present a limitation for the system performance. Hence, it is necessary to conduct the perturbative analysis and elucidate the system performance tolerance on the residual piston phase errors of the dynamic control system.

Figure 17(a) exhibits the variations of the performance metric GBPF with the increase of RMS residual piston phase errors. For both the cases of structuring OAM+2 and OAM+4 beams, calculations have been carried out and the results are illustrated by the blue and red curves, respectively. For each value of RMS phase errors, one hundred sets of residual phase errors are randomly generated, and the GBPF metric is calculated as the averaged value. As for the customization of the OAM beam with the TC of a larger absolute value, more significant descending trend of the GBPF with the increase of RMS phase errors can be observed. To ensure the <1% loss of GBPF metric, the RMS residual piston phase errors of the 30-element coherent laser array should be lower than $\lambda/40$ and $\lambda/50$ for structuring the OAM beams with the TCs of +2 and +4, respectively. Even for the dynamic control system that only compensates the phase errors with the RMS residual errors of $\lambda/12.5$, 90% of the ideal GBPF metric can be achieved for the customization of OAM+2 beams. When the RMS phase errors equal to $\lambda/40$, $\lambda/25$, $\lambda/20$, and $\lambda/10$, the averaged intensity distributions of the combined OAM+2 beams are shown in Figs. 17(b1)-17(b4), respectively, and Figs. 17(c1)-17(c4), in turn, depict the averaged intensity distributions of the combined OAM+4 beams. As the RMS residual phase errors increases, the wavefront of the laser array deviates the desirable helical structure more seriously, which leads to the more apparent distortion of intensity profile in the far-field. Consequently, the energy of the combined OAM beam gradually spreads from the main ring to the positions of the inner vortex core and the outer surrounding sidelobes.

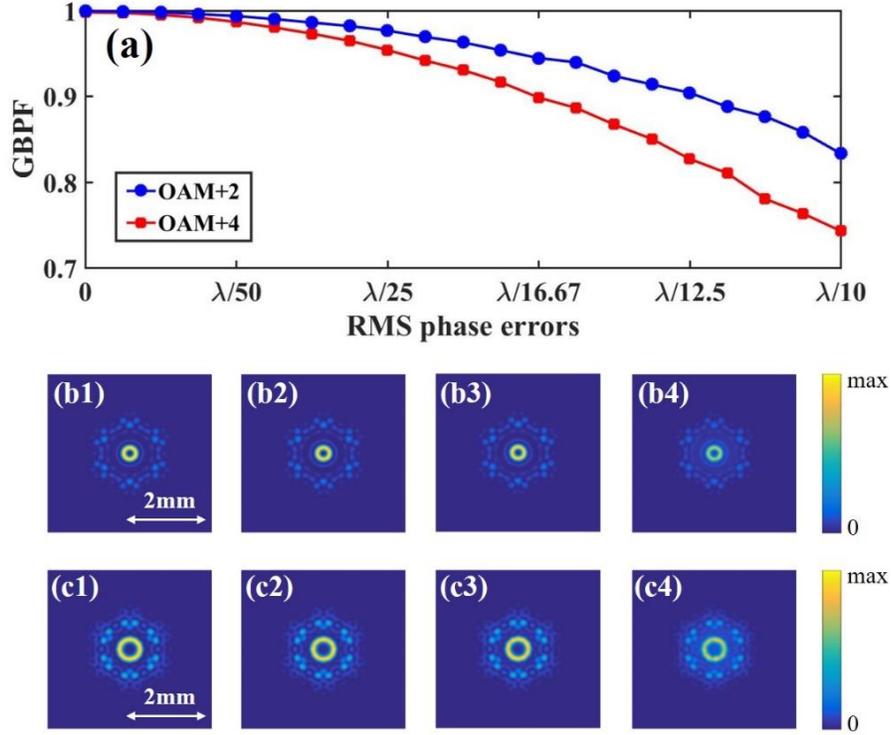

Fig. 17. Effect of piston phase errors on performance of combined OAM beams. (a) GBPF metric as a function of RMS residual phase errors for structuring OAM+2 and OAM+4 beams. (b1), (b2), (b3), and (b4) exhibit the averaged intensity distributions (calculated 100 times) of the combined OAM+2 beams when the RMS phase errors equal to $\lambda/40$, $\lambda/25$, $\lambda/20$, and $\lambda/10$, respectively. (c1), (c2), (c3), and (c4) are the averaged intensity distributions of the combined OAM+4 beams, which correspond to the RMS phase errors of $\lambda/40$, $\lambda/25$, $\lambda/20$, and $\lambda/10$, respectively.

Based on the analysis above, we have determined the effect of piston phase errors on the performance of combined OAM beams. Although we only consider the case of the laser array system that consists of 30 elements and discuss the OAM beams with the TCs of +2 and +4, the perturbative analysis for phase errors is general and can further be extended to the laser array systems that contains array elements of different numbers to tailor the OAM beams with a wide range of TCs. To design and construct a specific coherent fiber laser array system for structuring OAM beams with determined parameters, the perturbative analysis and corresponding results for the tolerance on residual piston phase errors could provide practical guidance on identifying the requirements for the dynamic control system.

**5.2 Tip-tilt wavefront control**

Owing to the mismatch of assembling and the dithering of fiber tails, tip-tilt wavefront errors are induced and should be treated. Similar with the piston phase errors, tip-tilt wavefront errors of array elements that cause far-field displacements of beamlets would also degrade the performance of the combined OAM beams. In our previous work, the necessity for compensating the tilt distortions in the laser array system that tailors OAM beams has been illustrated, whereas the rigorous analysis and quantitative identification of the system tolerance on the tilt distortions were not proposed due to the lack of appropriate performance metric for system evaluation [69]. In this work, the GBPF as the performance metric has been proposed, and therefore, investigating the effect of tip-tilt wavefront errors on the system performance and further determining the requirements for the dynamic control system in terms of tip-tilt wavefront control are highly required.

Before we discuss the results of perturbative analysis for the tip-tilt wavefront errors, the following assumptions should be clarified. Firstly, the goal of dynamic control system for tip-tilt wavefront control is to mitigate the tilt errors and ensure that each beamlet is pointing parallel to the optical axis, instead of applying active tilt optimization. It has been verified that employing tilt optimization can improve the similarity between the phase structure of the laser array with that of the desirable OAM beam [69]. However, in the laser array system for structuring OAM beams, the electro-optic PMs for applying piston phase control is of high rate (could reach gigahertz level), whereas the operation frequency of AFOC for modulating the tip-tilt wavefront is limited at the kilohertz level [46-48]. The current capacity of devices allows the real time compensation of tilt wavefront errors, whereas companying with the piston phase shifting to realize fast OAM mode switching is still challenging. Hence, the active tilt optimization that varies with the TC of combined OAM beam is not employed in the dynamic control system. The second assumption is related to the setting of tilt errors in the simulated environment. The AFOC for tilt wavefront control is capable of limiting the tilt angle of the beamlet within a specific range, thus utilizing the RMS value to characterize tilt errors is improper. Here we define the tilt errors as $\Delta\theta/\theta$, where $\Delta\theta$ represents the upper limit of the tilt deviation range, and $\theta = \lambda/(\pi w_0)$ denotes the divergence angle of each Gaussian beamlet [31]. For the $j$-th beamlet lying on the $n$-th radial subarray, the magnitude of tilt error $\Delta\theta_{j,n}$ is randomly generated within the range of $[0, \Delta\theta]$, thus the tilt angle $(\Delta\alpha_{j,n}, \Delta\beta_{j,n})$ in Eq. (2) is given by $(\Delta\theta_{j,n}\cos\Delta\Omega_{j,n}, \Delta\beta_{j,n}\sin\Delta\Omega_{j,n})$, where $\Delta\Omega_{j,n}$ is randomly generated within the range of $[0, 2\pi]$. According to our assumptions, the numerical model for the perturbative analysis of tilt errors can be constructed.

The effect of tip-tilt wavefront errors on the performance of combined OAM beams is analyzed based on the numerical model, and the results are illustrated in Fig. 18. Figure 18(a) depicts the GBPF metric as a function of the tilt errors $\Delta\theta/\theta$, and the blue and red curves account for the cases of structuring OAM+2 and OAM+4 beams, respectively. The calculated GBPF metric corresponding to each value of tilt errors is the averaged results of one hundred times simulations. The degradation of GBPF metric can be observed as the tilt errors become more serious, and one can observe that the descending trend for the case of structuring OAM+4 beams is more apparent, which is analogous to the effect of piston phase errors. To ensure the <1% loss of GBPF metric, the tilt errors of the 30-element coherent laser array should not exceed 0.9 and 0.55 for structuring OAM+2 and OAM+4 beams, respectively. As for the typical cases of tip-tilt wavefront errors when $\Delta\theta/\theta = 0.25$, $\Delta\theta/\theta = 0.5$, $\Delta\theta/\theta = 0.75$, and $\Delta\theta/\theta = 1$, the averaged intensity distributions of the combined OAM+2 beams are shown in Figs. 18(b1)-18(b4), respectively, and the averaged intensity distributions of the combined OAM+4 beams are in turn displayed in Figs. 18(c1)-18(c4). Comprehensively considering the loss of GBPF and averaged intensity profiles, we can identify that compared to the results of piston phase errors, the intensity distortions and purity degradation caused by the uncompensated tilt errors are relatively less. Moreover, the current technique of AFOC for tip-tilt wavefront errors compensation is capable of limiting the tilt errors within $\Delta\theta/\theta = 0.5$, indicating that the <1% loss of GBPF metric could be realized. Therefore, for the aspect of tip-tilt wavefront control, dynamic control system satisfies the requirements for structuring OAM beams with high performance.

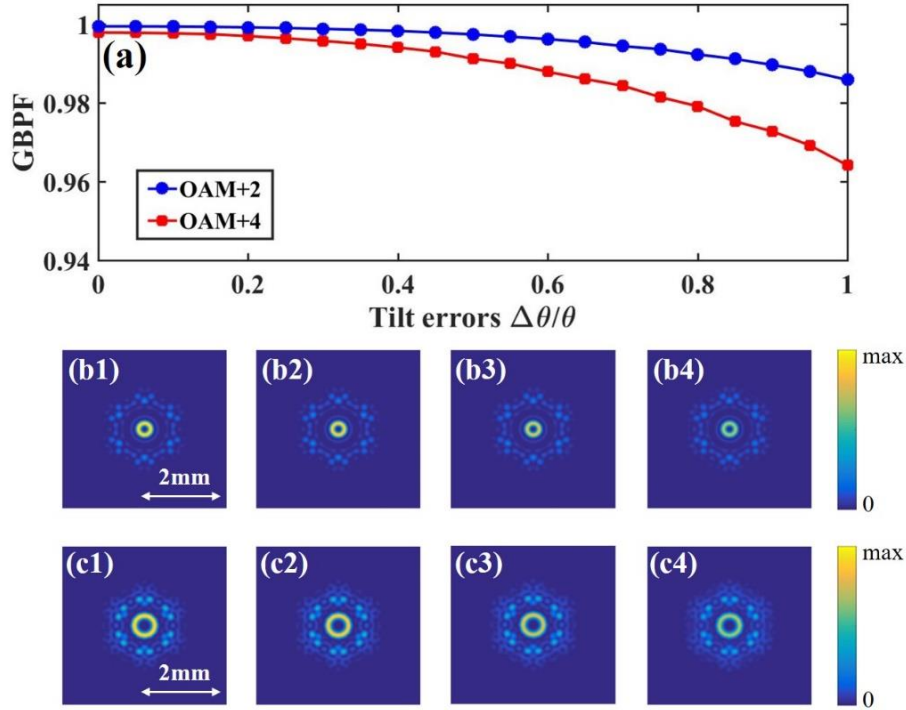

Fig. 18. Effect of tip-tilt wavefront errors on the performance of combined OAM beams. (a) GBPF metric as a function of tilt errors for structuring OAM+2 (blue) and OAM+4 beams (red). Averaged intensity distributions of combined OAM+2 beams for the cases of (b1) $\Delta\theta/\theta = 0.25$, (b2) $\Delta\theta/\theta = 0.5$, (b3) $\Delta\theta/\theta = 0.75$, and (b4) $\Delta\theta/\theta = 1$. Averaged intensity distributions of combined OAM+4 beams for the cases of (b1) $\Delta\theta/\theta = 0.25$, (b2) $\Delta\theta/\theta = 0.5$, (b3) $\Delta\theta/\theta = 0.75$, and (b4) $\Delta\theta/\theta = 1$.

## 6. Conclusion

In this work, the design considerations and performance evaluation of the coherent fiber laser array systems that tailor OAM beams have been comprehensively studied. We first introduce the model of laser array system, including the system architecture and the mathematical characterization of laser array. Then, the inappropriateness use of the existed criterions for evaluating the performance of coherently combined OAM beams has been elucidated, suggesting the necessity for the proposal of appropriate performance metric and evaluation method. Accordingly, we put forward the GBPF as the performance metric and introduce the general method for the measurement and calculation of the GBPF metric to evaluate the system performance.

Based on the GBPF metric we have proposed, we holistically analyze and identify the effect of the characteristics of the three main sections of the coherent fiber laser array system for structuring OAM beams (namely the high-power laser sources, emitting laser array configuration, and dynamic control system) on the system performance. According to the results of perturbative analysis for the high-power laser sources, we have found that the performance of combined OAM beams is not sensitive to the depolarization among beamlets when compared to the power variations. As a special case of power variations, the damage of array elements would cause the apparent degradation of the combined OAM beams, and we have noticed that the loss of the performance metric is related to the positions of damaged array elements.

In terms of the analysis for the emitting laser array configuration, the intuitive knowledge and routine insights of the conventional CBC systems for brightness enhancement have been extended when the customization of OAM beams is considered. Specifically, in spite of the advantage in energy concentration for the hexagonal arrangement, circular arrangement that ensures a higher OAM mode purity is preferred for structuring OAM beams with laser arrays. The filling of conformal aperture and subaperture is related to the combining efficiency, which is analogous to the perceptions of conventional CBC systems. Notably, different phenomena can be observed that the purity of the combined OAM beam with beam purification almost remains at the high level, and the optimized value of subaperture fill factor depends on the TC of the combined beam.

As for the perturbative analysis of the dynamic control system, piston and tilt wavefront distortions as the two main concerns for the performance degradation have been considered. The results indicate that when compared to the tilt errors, the effect of piston errors on the system performance is more serious, and the requirements for the dynamic control system are identified. Besides, we have also found that different from conventional CBC systems for brightness enhancement, the coherent fiber laser array systems for structuring OAM beams have a high tolerance on the tilt errors.

In conclusion, the tolerance of the laser array system for structuring OAM beams on the misalignments and errors is relatively higher for the case of structuring the OAM beam with the TC of lower absolute value. Among the perturbative factors that limit the performance of the system, the residual piston phase errors of the dynamic control system are the most important issue that should be treated for the further improvement of system performance. For one thing, our results and findings could provide valuable references on the requirements and optimization methods for the laser array system. For another thing, the GBPF metric, evaluation method, and system model are general and can be adaptively employed to analyze the laser array systems that customize OAM beams with versatile geometric arrangement, diverse numbers of array elements, and aiming at different applications. This work could be beneficial to guide the further development of the laser array based technical route for the controllable generation of high-power and fast switchable OAM beams.